\documentclass[preprintnumbers, prd, onecolumn, floatfix,preprintnumbers,
superscriptaddress, nofootinbib]{revtex4}
\usepackage{amssymb}
\usepackage{latexsym}
\usepackage{epsfig}
\usepackage{amsmath}
\usepackage{graphicx}
\usepackage{epsfig}
\usepackage{bm}
\usepackage{cancel}
\usepackage{amssymb}
\usepackage{float}
\usepackage{subfigure}
\usepackage{dcolumn}
\usepackage[colorlinks]{hyperref}
\usepackage[usenames,dvipsnames]{color}

\hypersetup{
     breaklinks=true,
    pdfstartview={FitH},        colorlinks=true,           linkcolor=blue,              citecolor=red,            filecolor=magenta,          urlcolor=blue,               anchorcolor=green,          linktocpage=true
}

\begin{document}

\title{Dynamics of an Interacting Barrow Holographic Dark Energy Model and
its Thermodynamic Implications}
\author{Abdulla Al Mamon}
\email{abdulla.physics@gmail.com}
\affiliation{Department of Physics, Vivekananda Satavarshiki Mahavidyalaya (affiliated to
the Vidyasagar University), Manikpara-721513, West Bengal, India}
\author{Andronikos Paliathanasis}
\email{anpaliat@phys.uoa.gr}
\affiliation{Institute of Systems Science, Durban University of Technology, PO Box 1334,
Durban 4000, South Africa}
%\affiliation{Instituto de Ciencias F\'{\i}sicas y Matem\'{a}ticas, Universidad Austral de
%Chile, Valdivia 5090000, Chile}
\author{Subhajit Saha}
\email{subhajit1729@gmail.com}
\affiliation{Department of Mathematics, Panihati Mahavidyalaya, Kolkata 700110, West
Bengal, India}

\begin{abstract}
In this paper, using Barrow entropy, we propose an interacting model of
Barrow holographic dark energy (BHDE). In particular, we study the evolution
of a spatially flat FLRW universe composed of pressureless dark matter and
BHDE that interact with each other through a well-motivated interaction
term. Considering the Hubble horizon as the IR cut-off, we then study the
evolutionary history of important cosmological parameters, particularly, the
density parameter, the equation of state parameter, and the deceleration
parameter in the BHDE model and find satisfactory behaviors in the model. We
perform a detailed study on the dynamics of the field equations by studying
the asymptotic behavior of the field equations, while we write the analytic
expression for the scale factor with the use of Laurent series. Finally, we
study the implications of gravitational thermodynamics in the interacting
BHDE model with the dynamical apparent horizon as the cosmological boundary.
In particular, we study the viability of the generalized second law by
assuming that the apparent horizon is endowed with Hawking temperature and
Barrow entropy.
\end{abstract}

\maketitle

%%%%%%%%%%%%%%%%%%%%%%%%%%%%%%%%%%%%%%%%%%%

%%%%%%%%%%%%%%%%%%%%%%%%%%%%%%%%%%%%%%%%%%%%%%%%%%%%%%%%%%%%%%%%%%%%%%%%%%%%

\section{Introduction}

%%%%%%%%%%%%%%%%%%%%%%%%%%%%%%%%%
Observational data from various probes \cite{acc1,acc2,acc3,acc4,acc5}
suggest that the expansion of the universe is accelerating at present. This
accelerated expansion is attributed to some exotic component with large
negative pressure called dark energy (DE) that comprises approximately 70\%
of the energy density of the universe. In addition, the second largest
component of our universe is the dark matter (DM), and the origin as well as
the true nature of these dark sectors (DE and DM) are absolutely unknown at
present. Different kinds of theoretical models have already been constructed
to interpret accelerating universe and some eminent reviews on this topic
can be found in \cite{de1,de2,de3}. However, the problem of the onset and
nature of this acceleration mechanism remains an open challenge of modern
cosmology.\newline

One interesting approach for the quantitative description of DE arises from
the holographic principle \cite{hp1,hp2,hp3,hp4,hp5,hp6}. Holographic dark
energy (HDE) leads to interesting cosmological scenarios, both at its simple
as well as at its extended versions, which mainly based on the use of
various horizons as the universe ``radius'' (see these Refs. \cite%
{hde1,hde2,hde3,hde4,hde5,hde6,hde7,hde8,hde9,hde10,hde11,hde12,hde13,hde14,hde15,hde16,hde17,hde18,hde19,hde20}
for more details about the models). Such HDE models are also in agreement
with observational data \cite%
{hdeo1,hdeo2,hdeo3,hdeo4,hdeo5,hdeo6,hdeo7,hdeo8}. Barrow holographic dark
energy (BHDE) is also an interesting alternative scenario for the
quantitative description of DE, originating from the usual holographic
principle \cite{hp1,hp2,hp3,hp4,hp5} and by applying the 
Barrow entropy \cite{barrow} instead of the usual Bekenstein-Hawking one 
\cite{Bekenstein1,Bekenstein2}. Recently, Saridakis \cite{Saridakis}
shown that the BHDE includes basic HDE as a sub-case in the limit where
Barrow entropy becomes the usual Bekenstein-Hawking one, but which in
general is a new scenario which reveals more richer and interesting
phenomenology. Very recently, Anagnostopoulos et al. \cite{Saridakis2} have
shown that the BHDE is in agreement with observational data, and it can
serve as a good candidate for the description of DE. On the other hand,
concerning various cosmological theories, the scenario where DE interacts
with DM has gained much attention in the current literature (for review, see 
\cite{intreview} and references therein). In fact, recently it has been
argued that the interacting model could be a promising candidate to resolve
the small value of the cosmological constant \cite{de2, intreview} and the
current tension on the local value of the Hubble constant \cite{h01,h02}.
Therefore, an interacting scenario seems promising and it might open some
new possibilities regarding the true nature of dark sectors in near future.%
\newline

Thus, following this motivation, in the present work, we propose an
interacting BHDE model in which the dark sectors (pressureless DM and BHDE)
of the universe interacts with each other through a general source term $Q$.
The basic properties and the physical motivations behind the choice of this $%
Q$ has been discussed in the next section. In particular, we consider a
spatially flat, homogeneous and isotropic spacetime as the underlying
geometry. We then study the behavior of different cosmological parameters
(e.g., the deceleration parameter, the density parameter of BHDE and the
equation of state parameter of BHDE) during the cosmic evolution by assuming
the Hubble horizon as the infrared (IR) cut-off. A suitable justification
for considering such a cut-off is provided in section \ref{sec2}.\newline

The asymptotic behavior of the field equation is studied by using the
Hubble-normalization parameters. The field equations admit two stationary
points where the one point describes a scaling solution while the second
stationary point describes the de Sitter universe. Moreover, for a specific
value of the parameters an exact singular solution it is determined, while
by using the singularity analysis we are able to write the analytic solution
of the model by using Laurent expansions around the initial singularity.%
\newline

Finally, we undertake a thermodynamic study of our interacting BHDE model.
We study the validity of the generalized second law (GSL) by assuming the
dynamical apparent horizon as the thermodynamic boundary. To meet our
purpose, we consider that the apparent horizon is endowed with Hawking
temperature and Barrow entropy.\newline

We organize the present work in the following way. In the next section, we
introduce the BHDE model proposed in \cite{Saridakis} with a general
interaction term between the dark components (BHDE and DM) of the universe
and also study its cosmological evolution. For completeness of our study, in
section \ref{secdyna}, we present an analysis by studying the dynamics of
the field equations and specifically its equilibrium points. Moreover, in
section \ref{secthermo}, we explore the thermodynamical properties of the
present model. Finally, in section \ref{conclusion} we draw our conclusions. 
\newline

Throughout the paper, $G$ is the Newton's gravitational constant and we have
used units where $\hbar = \kappa_B = c = 1$. As usual, the symbol dot
denotes derivative with respect to the cosmic time $t$ and a subscript zero
refers to value of the quantity evaluated at the current epoch. 
%%%%%%%%%%%%%%%%%%%%%%%%%%%%%%%%%%%%%%%%%%%%%%%%%%%%%%%%%%%%%%%%%%%%%%%%%%%%%%%%%%%%%%%%%%%%%%%

\section{The model}

\label{sec2} 
%%%%%%%%%%%%%%%%%%%%%%%%%%%%%%%%%%%%%%%%%%%%%%%%%%%%%%%%%%%%%%%%%%%
In this section, we describe in a nutshell the theoretical framework and the
cosmological scenario of an interacting BHDE model. Very recently, it was
shown by Barrow \cite{barrow} that the horizon entropy of a black hole may
be modified as 
\begin{equation}  \label{eq-barrow}
S_{B}=\left(\frac{A}{A_0}\right)^{1+\frac{\Delta}{2}},~~~0\leq \Delta \leq 1,
\end{equation}
where $A$ is the standard horizon area and $A_0$ indicates the Planck area.
In equation (\ref{eq-barrow}), the quantum deformation is quantified by the
new exponent $\Delta$. It is important to note here that the value $\Delta =
1$ corresponds to maximal deformation, while the value $\Delta=0$
corresponds to the simplest horizon structure, and in this case one can
recover the usual Bekenstein entropy \cite{Bekenstein1,Bekenstein2}. It is
important to note here that the entropy, as given in equation (\ref%
{eq-barrow}), resembles Tsallis non-extensive entropy \cite{te1,te2}, but
the involved physical principles and foundations are completely different. While standard HDE is given by the inequality $\rho_{D}L^{4} \le S$, where $L$ denotes the IR cutoff, and under the imposition $S \propto A \propto L^{2}$ \cite%
{hp1,hp2,hp3,hp4,hp5,hp6}, the use of Barrow entropy (\ref{eq-barrow}) will lead to
\begin{eqnarray}  \label{Trho}
\rho_D=C L^{\Delta-2},
\end{eqnarray}
where $C$ is an unknown parameter \cite{Saridakis}. The above
relation leads to some interesting results in the holographical and
cosmological setups \cite{Saridakis,Saridakis2}. It is notable that for the
special case $\Delta =0$, the above relation provides the usual HDE, i.e., $%
\rho_{D}\propto L^{-2}$. Therefore, the BHDE is indeed a more general
framework than the standard HDE scenario and hereafter, we focus on the
general case ($\Delta>0$), where the quantum deformation effects switch on.
If we consider the Hubble horizon ($H^{-1}$) as the IR cutoff ($L$), then
the energy density of BHDE is obtained as 
\begin{eqnarray}  \label{Hrho}
\rho_D=C H^{2-\Delta}.
\end{eqnarray}
%%%%%%%%%%%%%%%%
At this point, few comments on the choice of Hubble horizon as the IR cutoff
are in order. The Hubble horizon is undoubtedly the most natural length
scale in the context of Cosmology. In this regard, it is worthwhile to note
that different models of HDE have been studied in the literature with the
assumption of a wide range of IR cutoffs. Li \cite{hde1} observed that when
there is no interaction between DM and DE, the choice of future event
horizon as the IR cutoff gives the desired scenario of an accelerating
universe, while the particle horizon leads to a decelerating universe. On
the other hand, Hsu \cite{new2} demonstrated that the Hubble horizon leads
to an EoS of dust ($w=0$). However, when the interaction between DM and DE
is taken into account, the choice of Hubble horizon can not only produce an
accelerating universe but also solve the coincidence problem \cite{hde5,new3}%
. Thus, our choice of Hubble horizon as the IR cutoff in the present context
of interacting BHDE is quite justified. \newline

Let us consider a spatially flat, homogeneous and isotropic Friedmann-Lema%
\^{\i}tre-Robertson-Walker (FLRW) universe endowed with the standard metric 
\begin{equation}
ds^{2}=-dt^{2}+a^{2}(t)\delta _{ij}dx^{i}dx^{j}.
\end{equation}%
We further assume that the Universe is filled with pressureless DM and BHDE.
Then the corresponding Friedmann equation and the acceleration equation are
obtained as 
\begin{eqnarray}  \label{frd}
H^{2} &=&(8\pi G/3)\left( \rho _{m}+\rho _{D}\right) ,  \label{frd1} \\
\dot{H} &=&-4\pi G(\rho _{m}+\rho _{D}+p_{m}+p_{D}),  \label{frd2}
\end{eqnarray}%
where, $H(t)=\frac{\dot{a}(t)}{a(t)}$ is the Hubble parameter and $a(t)$ is
the scale factor of the universe. Parameters $\rho _{m}$, $p_{m}$ correspond
to the energy density and the pressure of DM respectively, while $\rho _{D}$%
, $p_{D}$ correspond to the energy density and the pressure of BHDE
respectively. The conservation of the total energy-momentum tensor leads to
the continuity equation 
\begin{equation}
\dot{\rho}_{m}+\dot{\rho}_{D}+3H(\rho _{m}+\rho _{D}+p_{m}+p_{D})=0.
\label{emce}
\end{equation}%
The fractional energy density parameters of BHDE ($\Omega _{D}$) and DM ($%
\Omega _{m}$) are, respectively, given by 
\begin{eqnarray}
&&\Omega _{D}=\frac{\rho _{D}}{\rho _{c}}=(8\pi G/3)CH^{-\Delta },  \label{3}
\\
&&\Omega _{m}=\frac{\rho _{m}}{\rho _{c}},
\end{eqnarray}%
where, $\rho _{c}=(3/8\pi G)H^{2}$ is the critical energy density. Now,
equation (\ref{frd}) can be rewritten as 
\begin{equation*}
\Omega _{m}+\Omega _{D}=1.
\end{equation*}%
%
%%%%%%%%%%%%%%%%%%%%%%%%%%%%%%%%%%%%%%%%%%%%%%%%%%%%%%%%%%%%%%%%%%%
In addition, we assume that the dark fluids (BHDE and pressureless DM) of
the universe exchange energy through an interaction term $Q$. Therefore, we can write the conservation equations $T_{~~~;\nu }^{\mu \nu}=0$ both for DM $^{\left( DM\right) }T^{\mu \nu }~$ and BHDE $^{\left( DE\right) }T^{\mu \nu }$, $~^{\left( DM\right)}T_{~;\nu }^{\mu \nu }~+^{DE}T_{~;\nu }^{\mu \nu }=0$ in the
following coupled form
\begin{eqnarray}
&&\dot{\rho}_{m}+3H\rho _{m}=Q,  \label{conm} \\
&&\dot{\rho}_{D}+3H(1+\omega _{D})\rho _{D}=-Q,  \label{conD}
\end{eqnarray}%
%
%%%%%%%%%%%%%%%%%%%%
where, $\omega _{i}=\frac{p_{i}}{\rho _{i}}$ is the \textit{equation of state%
} (EoS) parameter of the corresponding fluid sector. In the above equations, 
$Q$ represents the rate of energy density transfer, where (A) Energy
transfer is from BHDE $\rightarrow $ DM, if $Q>0$; (B) Energy transfer is
from DM $\rightarrow $ BHDE, if $Q<0$.\newline

Hence, once the evolution of the energy densities $\rho _{m}$ and $\rho _{D}$
are determined either numerically or analytically for some given interaction
term $Q$, the expansion rate of the universe can be obtained and the
modified cosmological parameters can be described in terms of their
evolution with time. If we observe the energy conservation equations (\ref%
{conm}) and (\ref{conD}), the interaction between BHDE and DM must be a
function of the energy densities multiplied by a quantity having units of
the inverse of time which has the natural choice as the Hubble parameter $H$%
. Hence $Q$ could be expressed phenomenologically in any arbitrary forms,
for example, $Q\propto H\rho $ with assumptions of $\rho =\rho _{m}$, $\rho
=\rho _{D}$ and $\rho =\rho _{m}+\rho _{D}$ are more popular in this
context. Additionally, there are many proposed interactions in the
literature to study the dynamics of the universe and for review, one can
look into \cite{intreview} and references therein. Inspired by these facts
and also for mathematical simplicity, in the present work, we assume that $Q$
is a linear combination of the energy densities given as \cite%
{int1,int2,hde5,int4,int5,int6} 
\begin{equation}
Q=3H(b_{1}^{2}\rho _{m}+b_{2}^{2}\rho _{D}),  \label{eq-ans}
\end{equation}%
%
%%%%%%%%%%%%%%%%%%%
where, $b_{1}^{2}$ and $b_{2}^{2}$ are dimensionless constants. From the
observational point of view, the values of $b_{1}^{2}$ and $b_{2}^{2}$ are
very small ($<<1$) \cite{intobs}. Recently, Mamon et al. \cite{int6} studied
the cosmological and thermodynamical consequences of Tsallis holographic
dark energy with this choice of interaction (\ref{eq-ans}) and found it can
bring new features to cosmology. They have also showed that the general form
of $Q$, given by equation (\ref{eq-ans}), covers a wide range of other
well-known interacting models for some specific choices of $b_{1}^{2}$ and $%
b_{2}^{2}$. Furthermore, they justified the choice of this $Q$ using the
Teleparallel Gravity, based in the Weitzenb\"{o}ck spacetime (for details,
see \cite{int6}). Following Ref. \cite{int6}, in this work, we also focus on
the positive values of the coupling constants $b_{1}^{2}$ and $b_{2}^{2}$.
As a result, $Q$ becomes positive (and hence the energy transfers from BHDE
to DM) which is well consistent with the validity of the second law of
thermodynamics and Le Chatelier-Braun principle \cite{int2}. The simplicity
of the functional form of $Q$ (as given in equation (\ref{eq-ans})), makes
it very attractive to study. Clearly, equations (\ref{conm}) and (\ref{conD}%
) offer a new dynamics of the universe with this choice $Q$. Hence, such a
consideration might be useful and deserves further investigation in the
present context. For a detailed discussion on interacting models we refer
the reader in \cite{inan1,inan2}, while some recent cosmological constraints
on interacting models can be found for instance \cite{sp1,sp2,sp3}.\newline

Now, taking the time derivative of equation (\ref{frd}) along with combining
the result with equations (\ref{conm}) and (\ref{conD}), one can easily
obtain 
\begin{eqnarray}
\frac{\dot{H}}{H^{2}} &=&-\frac{3}{2}\Omega _{D}\left( 1+\omega _{D}+\frac{%
\rho _{m}}{\rho _{D}}\right)  \notag  \label{7} \\
&=&-\frac{3}{2}(1+\omega _{D}\Omega _{D}).
\end{eqnarray}%
%
%
%%%%%%%%%%%%%%%%%%%%%%%%%%%%%%%%%%%%%%%%%%%%%%%%%%%%%%%%%%%%%%%%%%%%%%%%%%%%%%%%%%%
Similarly, taking the time derivative of equation (\ref{Hrho}) and by using
equations (\ref{conD}) and (\ref{7}), we get 
\begin{equation}
\omega _{D}=\frac{2b_{1}^{2}(\Omega _{d}-1)-(2b_{2}^{2}+\Delta )\Omega _{D}}{%
\Omega _{D}(2-(2-\Delta )\Omega _{D})}.  \label{w1}
\end{equation}%
Now, the equation of motion for the BHDE density parameter $\Omega _{D}$ can
be evaluated by differentiating equation (\ref{3}) with respect to the
cosmic time and using equations (\ref{7}) and (\ref{w1}). Therefore, we
reach at %%%%%%%%%%%%%%%%%%%%%%%%%%%%%%%%%%%%%%%%%%%%%%%%%%%%%%%%%%%%%
\begin{equation}
\Omega _{D}^{\prime }=\frac{d\Omega _{D}}{d(\ln a)}=\frac{3\Delta \Omega
_{D}[1+b_{1}^{2}(\Omega _{D}-1)-(1+b_{2}^{2})\Omega _{D}]}{(2-(2-\Delta
)\Omega _{D})}.  \label{Omega}
\end{equation}%
%
%
%%%%%%%%%%%%%%%%%%%%%%%%%%%%%%%%%%%%%%%%%%%%%%%%%%%%%%%%%%%%%%%%
The deceleration parameter is defined as 
\begin{equation*}
q=-\frac{\ddot{a}}{aH^{2}}=-1-\frac{\dot{H}}{H^{2}},
\end{equation*}%
which finally leads to 
\begin{equation*}
q=-\frac{\ddot{a}}{aH^{2}}=\frac{1+3b_{1}^{2}(\Omega
_{D}-1)-(1+3b_{2}^{2}+\Delta )\Omega _{D}}{(2-(2-\Delta )\Omega _{D})}.
\end{equation*}%
%
%
%%%%%%%%%%%%%%%%%%%%%%%%%%%%%%%%%%%%%%%%%%%%%%%%%%%%%%%%
The total EoS parameter is also evaluated as 
\begin{equation}
\omega _{tot}=-1-\frac{2{\dot{H}}}{3H^{2}}=-\frac{1}{3}+\frac{2q}{3}.
\label{eq-totw}
\end{equation}%
As is well known, $\omega _{tot}<-\frac{1}{3}$ is require to accelerate the
expansion of our universe.\newline
%%%%%%%%%%%%%%%%%%%%%%%%%%%%%%%%%%%%%%%%%%%%%%%%%%%%%%%%%%%%%%
%%%%%%%%%%%%%%%%%%%%%%%%%%%%%%%%

For completeness, in the next section, we shall try to solve the field
equations and determine exact and analytic solutions. 
%%%%%%%%%%%%%%%%%%%%%%%%%%%%%%%%%%%%%%%%%%%%%%

\section{Asymptotic behavior of the dynamics}

\label{secdyna}

We proceed by studying the asymptotic behavior of the gravitational field
equaitons as also the existence of exact solutions for the field equations.
With the use of the dimensionless variables $\Omega _{m},~\Omega _{D}$ the
field equations reduce to the one-dimensional first-order ordinary
differential equation (\ref{Omega}). Equation (\ref{Omega}) is a nonlinear
equation which can not be integrated by using closed-form functions. Hence
we proceed its analysis by studying the dynamics of the equation and
specifically its equilibrium points \cite{epo1,epo2,epo3,epo4}.\newline

The right hand side of equation (\ref{Omega}) vanishes at the two points $%
P_{1}:\Omega _{D}=0$ and $P_{2}:\Omega _{D}=\frac{b_{1}^{2}-1}{%
b_{1}^{2}-b_{2}^{2}-1}$. \ The stationary point $P_{1}$ describes an exact
solution where only the matter source $\Omega _{m}$ contributes in the
cosmological solution, and the parameter for the equation of state for the
effective fluid is $\omega_{tot}\left( P_{1}\right) =-b_{1}^{2}$. On the
other hand, the physical solution at the stationary point $P_{2}$ describes
a universe where the two fluid source contributes, when $b_{1}^{2}\neq 1$,
while the point is physically accepted when $\left\{ b_{2}=0,~\left\vert
b_{1}\right\vert \neq 1\right\} $ and $\left\{ b_{2}\neq 0,~\left\vert
b_{1}\right\vert \leq \sqrt{1+\frac{b_{2}^{2}}{2}}\right\} $. \ Moreover,
the parameter for the equation of state for the effective fluid is $%
\omega_{tot}\left( P_{2}\right) =-1$ which means that the effective fluid
mimic the cosmological constant.\newline

In order to study the stability of the stationary points we write the
linearized system around the point and we determine the eigenvalue of the
equation at the stationary points. At $P_{1}$, the eigenvalue is $e\left(
P_{1}\right) =-\frac{3}{2}\left( b_{1}^{2}-1\right) \Delta $, while at $P_{2}
$ the eigenvalue is derived $e\left( P_{2}\right) =\frac{3\left(
b_{1}^{2}-1\right) \left( b_{1}^{2}-b_{2}^{2}-1\right) \Delta }{\Delta
\left( b_{1}^{2}-1\right) -2b_{2}^{2}}$. Recall that $0<\Delta <{1},$ from
where we infer that $P_{1}$ is an attractor when $\left\vert
b_{1}\right\vert >1$ and a saddle point for $\left\vert
b_{1}\right\vert <1$, while point~$P_{2}$ is an attractor for
arbitrary $b_{2}$ and $\left\vert b_{1}\right\vert <1$.\newline

\begin{figure}[th]
\begin{center}
\includegraphics[scale=0.55]{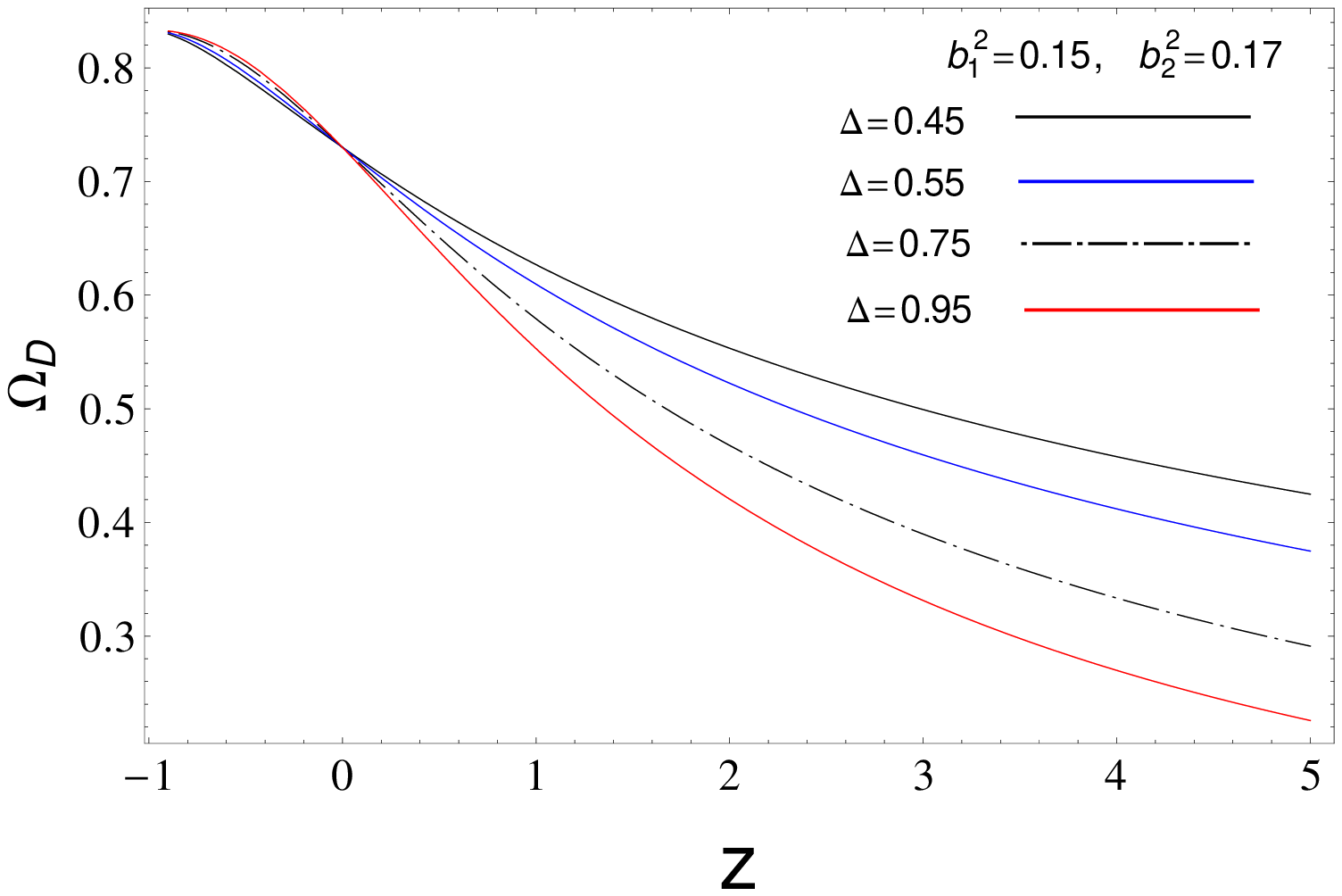}\\[0pt]
\includegraphics[scale=0.55]{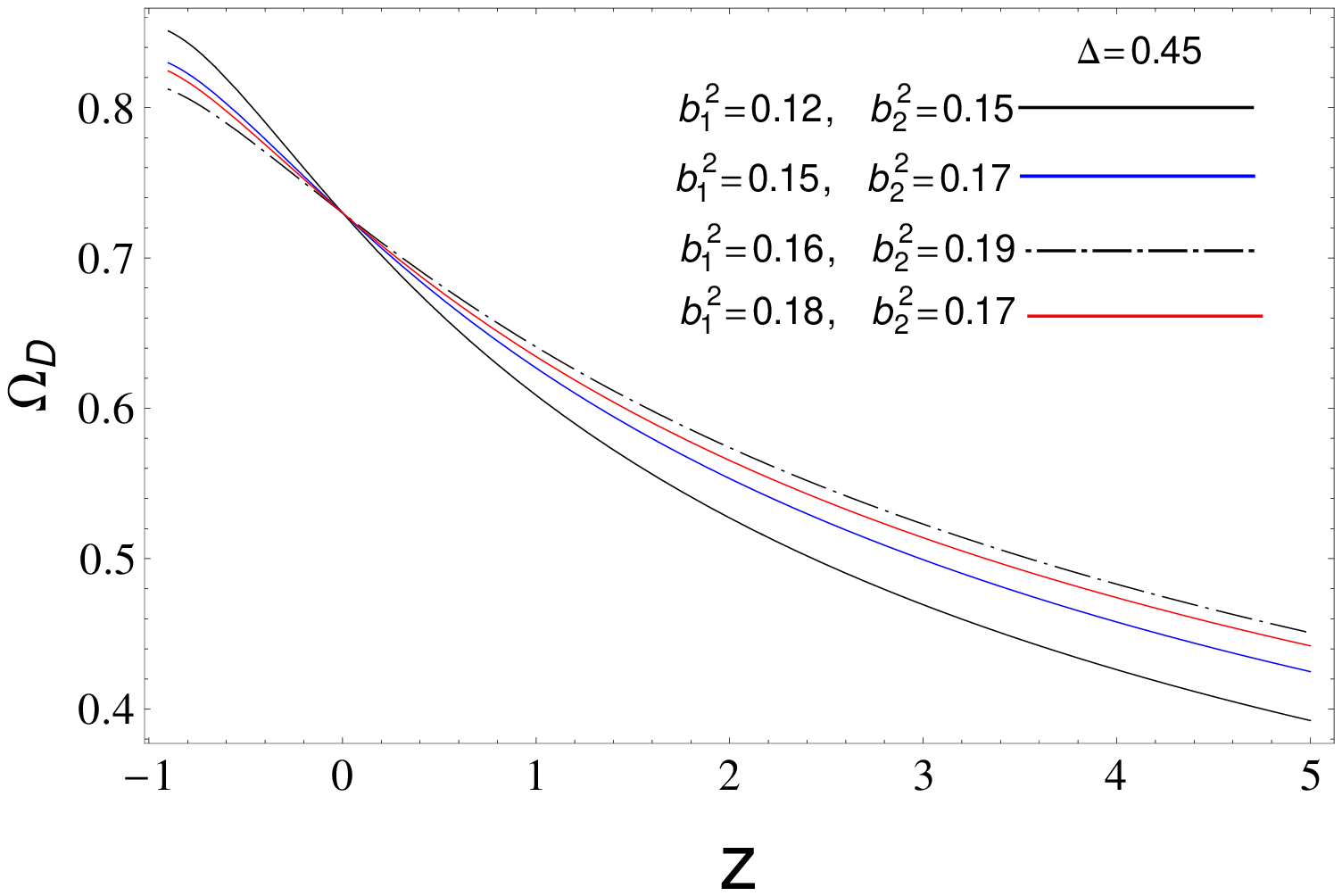}
\end{center}
\caption{The evolution of $\Omega _{D}$, as a function of $z$, is shown for
the present model considering $\Omega _{D0}=0.73$ and different values of $%
\Delta $, $b_{1}^{2}$ and $b_{2}^{2}$, as mentioned in each panel.}
\label{fo1}
\end{figure}

\begin{figure}[th]
\begin{center}
\includegraphics[scale=0.55]{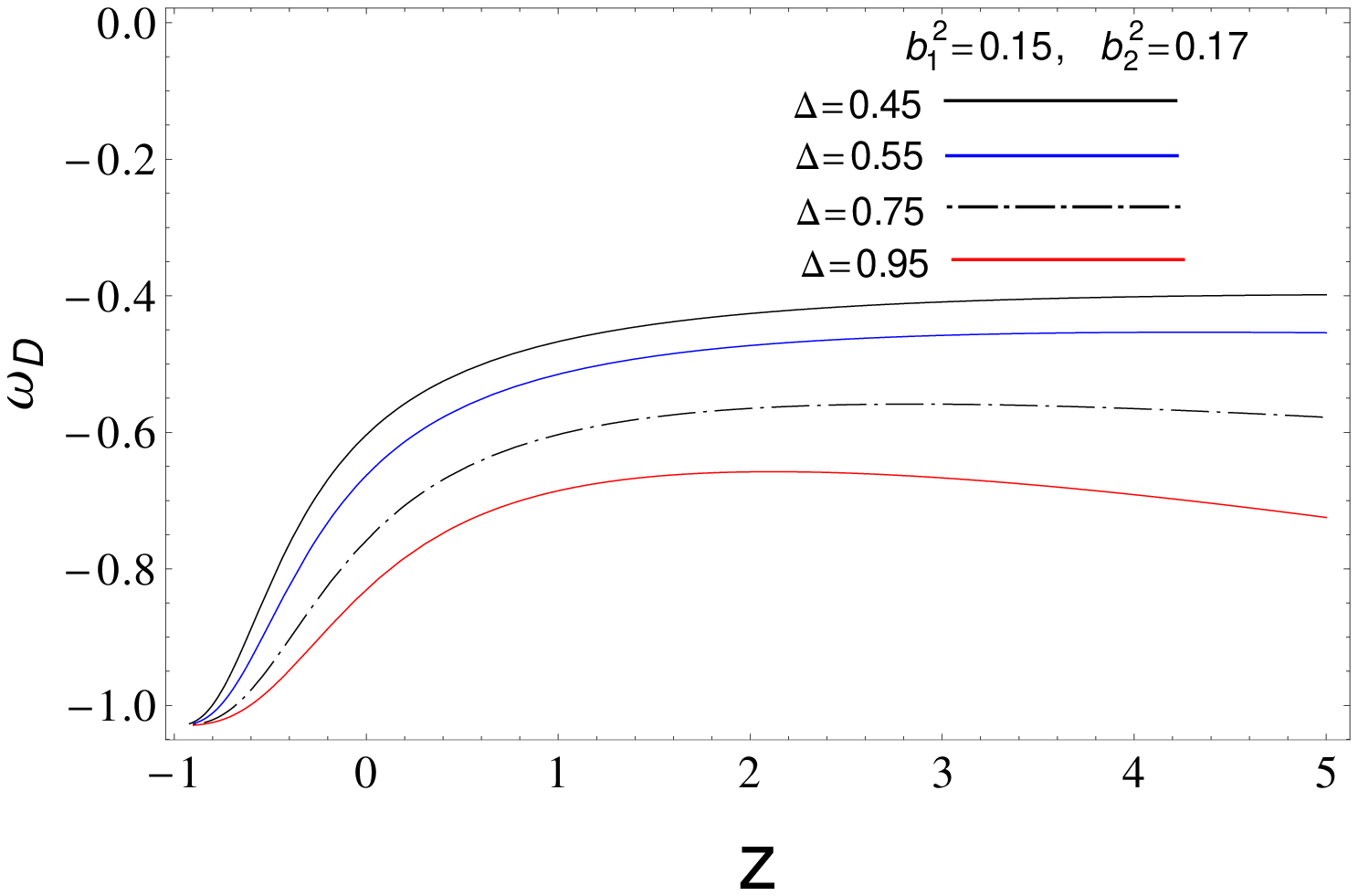}\\[0pt]
\includegraphics[scale=0.55]{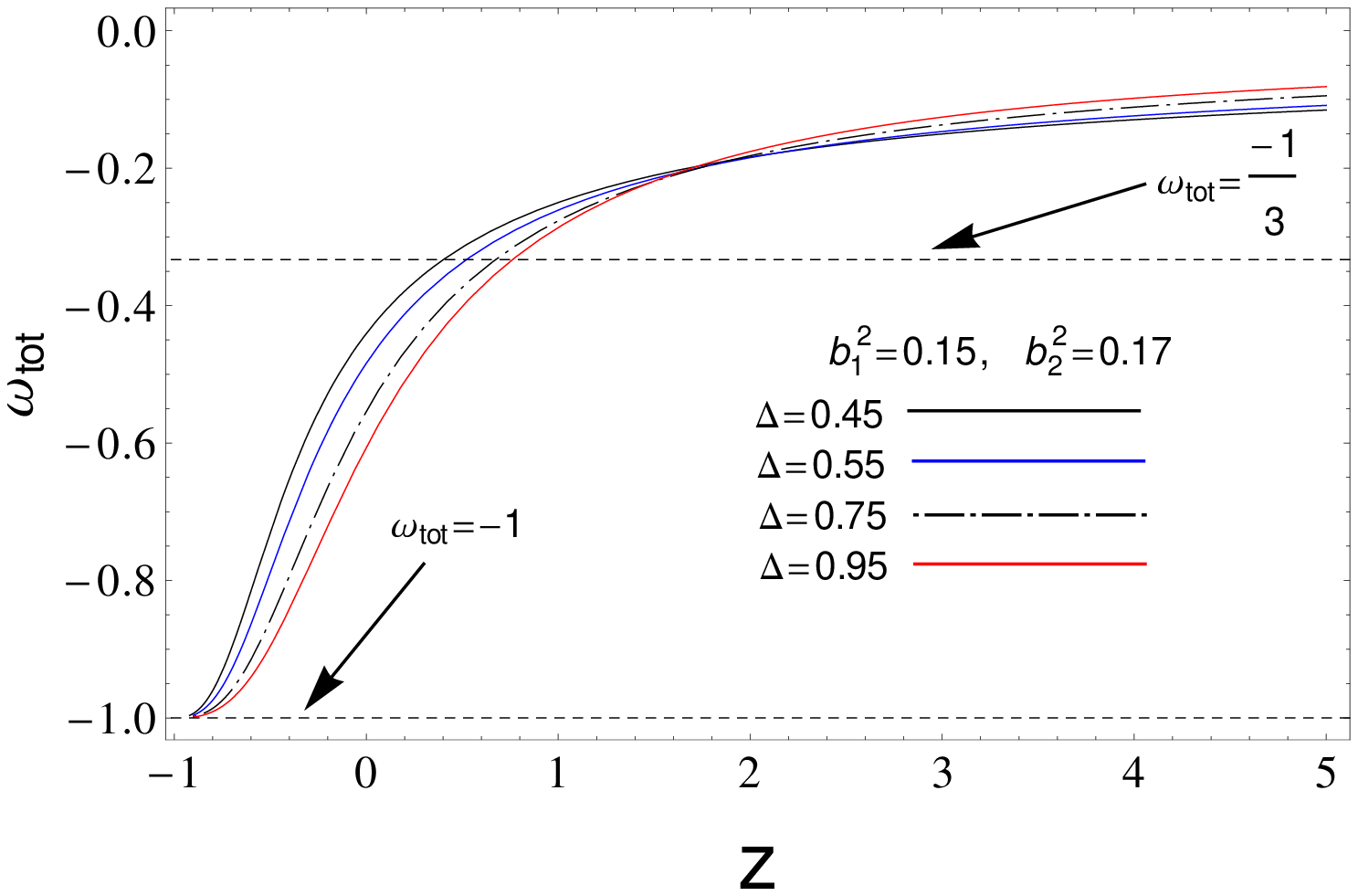}
\end{center}
\caption{The evolutions of $\protect\omega _{D}$ vs. $z$ (upper panel) and $%
\protect\omega _{tot}$ vs. $z$ (lower panel), are shown for $\Omega
_{D0}=0.73$, $b_{1}^{2}=0.15$, $b_{2}^{2}=0.17$ and different values of $%
\Delta $, as mentioned in each panel.}
\label{fo2}
\end{figure}

\begin{figure}[th]
\begin{center}
\includegraphics[scale=0.55]{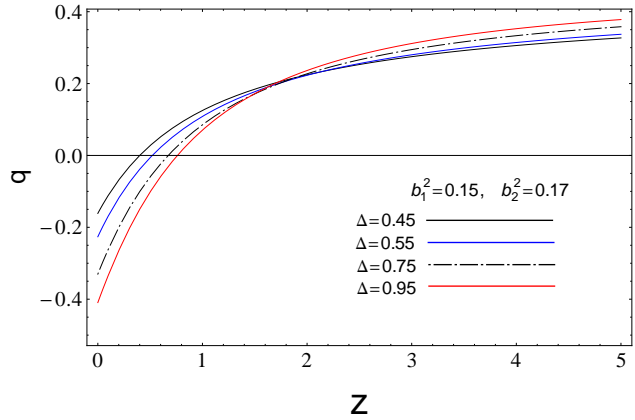}
\end{center}
\caption{The evolution of $q$, as a function of $z$, is shown for our model
considering $\Omega _{D0}=0.73$, $b_{1}^{2}=0.15$, $b_{2}^{2}=0.17$ and
different values of $\Delta $, as mentioned in each panel.}
\label{fo3}
\end{figure}

We plot the evolutionary trajectories for different cases of new Barrow
exponent $\Delta ~$in which $\left\vert b_{1}\right\vert <1$ which means
that the future attractor is point $P_{2}$.. Figure \ref{fo1} shows the
evolution of the BHDE density parameter $\Omega _{D}$ as a function of the
redshift parameter $z$. From this figure, it is evident that $\Omega _{D}$
increases monotonically to unity as the universe evolves to $z\rightarrow -1$%
. Next, we have shown the evolutions of the EoS parameter $\omega _{D}$ and
the total EoS parameter $\omega _{tot}$ for the present model by considering
different values of $\Delta $. The plot of $\omega _{D}$ versus redshift $z$
is shown in the upper panel of figure \ref{fo2}, while the corresponding
plot of $\omega _{tot}$ is shown in the lower panel of figure \ref{fo2}.
Interestingly, we observed that for different values of $\Delta $, the EoS
parameter $\omega _{D}$ lies in the quintessence regime ($\omega _{D}>-1$)
at the present epoch, however it enters in the phantom regime ($\omega
_{D}<-1$) in the far future (i.e., $z\rightarrow -1$). On the other hand, we
also observed from the lower panel of figure \ref{fo2} that the total EoS
parameter $\omega _{tot}$ was very close to zero at high redshift and
attains some negative value in between $-1$ to $-\frac{1}{3}$ at low
redshift and further settles to a value very close to $-1$ in the far
future. Moreover, the evolution of $q$ has been plotted in figure \ref{fo3}.
As we observed from figure \ref{fo3}, the interacting BHDE model can
describe the universe history very well, with the sequence of an early
matter dominated and late-time DE dominated eras. Additionally, the
transition redshift $z_{t}$ (i.e., $q(z_{t})=0$) occurs within the interval $%
0.5<z_{t}<1$, which are in good compatibility with different recent studies
(see Refs. \cite{zt1,zt2,zt3,zt4,zt5,zt6,zt7,zt8} for more details about the
models and cosmological datasets used). It has also been observed that the
parameter $z_{t}$ depends on the values of $\Delta $ in such a way that, as $%
\Delta $ increases, the parameter $z_{t}$ also increases.\newline
%%%%%%%%%%%%%%%%%%%%%%%%%%%%%%%%%%%%%%%%%%%%%%
%%%%%%%%%%%%%%%%%%%%%%%%%%%%%%%%%%%%%%%%%%%%%%
%%%%%%%%%%%%%%%%%%%%%%%%%%%%%%%%%%%%%%%%%%%%%%
%%%%%%%%%%%%%%%%%%%%%%%%%%%%%%%%%%%%%%%%%%%%%%
%%%%%%%%%%%%%%%%%%%%%%%%%%%%%%%%%%%%%%%%%%%%%%
%%%%%%%%%%%%%%%%%%%%%%%%%%%%%%%%%%%%%%%%%%%%%%%%%%%%%%%%%%%%%%%

\subsection{Exact and analytic solutions}

%%%%%%%%%%%%%%%%%%%%%%%%%%%%%%%%%%%%%%%%%%%%%%%%%%%%%%%%%%%%%%%%%%

Consider now the second Friedmann's equation which can be written in the
equivalent form 
\begin{eqnarray}
&&\left( \left( 2-\Delta \right) \Omega _{D0}-2H^{\Delta }\right) \dot{H} \\
+3\left( 1-b_{1}^{2}+b_{2}^{2}\right) \Omega _{D0}H^{2}-3\left(
1-b_{1}^{2}\right) H^{2+\Delta } &=&0  \notag  \label{sf1}
\end{eqnarray}%
where $\Omega _{D0}=(8\pi G/3)C$. The latter equation when $\Delta =\frac{%
2b_{2}}{b_{1}^{2}-1}$ admits the special exact solution $H\left( t\right) =%
\frac{2}{3\left( 1-b_{1}^{2}\right) \left( t-t_{0}\right) }$. The
latter exact solution describes the epoch when the universe is dominated by
an ideal gas with equation of state parameter $\omega _{tot}=-b_{1}^{2}$%
,~that is, it describes the solution at point $P_{1}$.%
\newline

Thus for arbitrary parameter $\Delta $, we observe that \ that the singular
behaviour $H\left( t\right) =\frac{2}{3\left( 1-b_{1}^{2}\right) \left(
t-t_{0}\right) }$ it is not an exact solution but describes the leading
terms of the second Friedmann's equation near the singularity $t-t_{0}=0$.
Consequently, the singularity analysis can be applied in order to determine
the analytic solution of the field equations. Singularity analysis is a powerful 
method for the determination of analytic solutions for differential equations, 
as also for the study of the integrability property for a given system.
Nowadays, the singularity analysis it is summarized in the so-called
Ablowitz, Ramani and Segur algorithm \cite{Abl1,Abl2,Abl3}, known also as
ARS algorithm. The latter method provides necessary information if a given
differential equation passes the Painlev\'{e} test and consequently if the
solution of the differential equation can be written as a Laurent expansion
around a movable singularity. This method has been widely applied in
gravitational studies, for instance see \cite%
{miritzis,cots1,cots2,bun2,cots,sinFR,sinFR2,sinFT,sinr02} and references
therein.\newline

There are three main steps for the ARS algorithm. The first step is the
determination of the leading-order behaviour, which we have already found it
in our approach. The second step has to do with the position of the
resonances. We replace 
\begin{equation}
H\left( t\right) =\frac{2}{3\left( 1-b_{1}^{2}\right) \left( t-t_{0}\right) }%
+\varepsilon t^{-1+S},
\end{equation}%
in equation (\ref{sf1}) and we linearize around $\varepsilon \rightarrow 0$.
From the leading-order terms we end with the algebraic equation for the
resonance $S+1=0$, that is $S=-1$, which indicates that the singularity is
movable. Because the differential equation is of first order we do not have
to continue our analysis, however for completeness on the presentation we
proceed with the third-step of the ARS algorithm, the consistency test.%
\newline

For the consistency test we select $\Delta =\frac{1}{2}$ and we replace $%
H\rightarrow \mathcal{H}^{2}$. Hence, we end with the analytic solution
expressed in the Laurent Series%
\begin{eqnarray}
\mathcal{H}\left( t\right) =H_{0}\left( t-t_{0}\right) ^{-\frac{1}{2}%
}+H_{1}+H_{2}\left( t-t_{0}\right) ^{\frac{1}{2}} \\
+H_{3}\left( t-t_{0}\right)+...+H_{N}\left( t-t_{0}\right) ^{-\frac{1}{2}+%
\frac{N}{2}}  \notag
\end{eqnarray}%
where 
\begin{equation*}
H_{0}=\sqrt{\frac{2}{3\left( 1-b_{1}^{2}\right) }}~,~H_{1}=\frac{\left(
b_{1}^{2}-1-4b_{2}^{2}\right) }{12\left( b_{1}^{2}-1\right) }\Omega _{D0}~,
\end{equation*}%
\begin{equation*}
H_{2}=-\frac{\left( b_{1}^{2}-1-4b_{2}^{2}\right) \left( 5\left(
b_{1}^{2}-1\right) -2b_{2}^{2}\right) }{32\sqrt{6}\sqrt{1-b_{1}^{2}}\left(
b_{1}^{2}-1\right) }~,~...~.
\end{equation*}%
We remark that the only integration constant is the location of the
singularity $t_{0}$. \ Finally, equation (\ref{sf1}) possess the Painlev\'{e}
property and it is integrable in terms of the singularity analysis. 
%%%%%%%%%%%%%%%%%%%%%%%%%%%%%%%%%%%%%%%%%%%%%%%%%%%%%%%%%%%%%%%%%%%%%

\section{Thermodynamic Implications of Interacting BHDE}

\label{secthermo} 
%%%%%%%%%%%%%%%%%%%%%%%%%%%%%%%%%%%%%%%%%%%%%%%%%%%%%%%%%%%%%%
We shall now proceed to study the thermodynamic implications of the
interacting BHDE proposed in this paper. To meet our purpose, we wish to
consider the dynamical apparent horizon of our homogeneous and isotropic
FLRW universe. Then, we shall investigate the GSL by evaluating the first
order entropy variation for the physical system bounded by the apparent
horizon in the framework of interacting BHDE. This sort of thermodynamic
study was initiated by Wang et al. \cite{Wang1} and later extended by Saha
and Chakraborty \cite{Saha1,Saha2,Chakraborty1}. It must be noted that the
GSL in these models were studied by assuming that the apparent horizon is
ensowed with the Bekenstein entropy and the Hawking temperature. Very
recently, the GSL at the dynamical apparent horizon was studied with the
Viaggiu entropy \cite{Saha3,Mamon1} which have shown some promising results.
Although there exist many horizons in Cosmology, but the most relevant one
in this context is the dynamical apparent horizon which is a marginally
trapped surface with vanishing expansion given by \cite%
{Hayward1,Hayward2,Bak1} 
\begin{equation}
R_A = \frac{1}{\sqrt{H^2+\frac{\kappa}{a^2}}},
\end{equation}
where $k$ is the spatial curvature which we shall set to zero, consistent
with our assumption of a spatially flat universe.\newline

At this juncture, it is worth mentioning that the standard operating
procedure for GSL study is to determine the sign of the first order entropy
variation of the apparent horizon plus the first order entropy variation of
the fluid contained within it. The GSL will be satisfied if this sum is
nondecreasing.\newline

The first step in this direction is to employ the first law of
thermodynamics (FLT) which will provide us with the first order entropy
variation of the fluid. In mathematical terms, the FLT is stated as 
\begin{equation}
TdS=dE+pdV,
\end{equation}
where, $T$ and $S$ are the temperature and the entropy of the fluid
respectively, $V=\frac{4}{3}\pi R_{A}^{3}$ is the volume of the fluid
bounded by the horizon, $E=V\rho$ is the internal energy of the fluid,
evaluated at the dynamical apparent horizon, and $\rho$ and $p$ are the
energy density and the pressure of the fluid respectively.\newline

Thus, the change in entropy of the matter and that of the dark energy become 
\begin{eqnarray}
dS_m &=& \frac{1}{T}\left(p_m dV + dE_m\right) ,  \label{dsm} \\
dS_D &=& \frac{1}{T}\left(p_D dV + dE_D\right) .  \label{dsd}
\end{eqnarray}
Note that the temperature $T$ has been kept the same in the above equations
due to the establishment of thermal equilibrium amongst different cosmic
fluids. Now, dividing equations (\ref{dsm}) and (\ref{dsd}) by $dt$ both
sides, we obtain 
\begin{eqnarray}
\dot{S}_{m} &=& \frac{1}{T}\left(p_m 4\pi R_{A}^{2} \dot{R}_{A} + \dot{E}%
_{m}\right) ,  \label{smd} \\
\dot{S}_{D} &=& \frac{1}{T}\left(p_D 4\pi R_{A}^{2} \dot{R}_{A} + \dot{E}%
_{D}\right) .  \label{sdd}
\end{eqnarray}
In the above equations, $\dot{R}_A = -\dot{H}/{H^2} = -\dot{H}R_{A}^{2}$.%
\newline

Finally, plugging in the time derivatives of 
\begin{eqnarray}
E_D = \frac{4}{3}\pi R_{A}^{3} \rho_{D}, \\
E_m = \frac{4}{3}\pi R_{A}^{3} \rho_{m}
\end{eqnarray}
into equations (\ref{smd}) and (\ref{sdd}) and using equation (\ref{emce}),
we arrive at the first order entropy variations of the matter and dark
energy, respectively, as \cite{Saridakis3} 
\begin{eqnarray}
\dot{S}_{m} &=& \frac{1}{T}(1+\omega_m)\rho_{m} 4\pi R_{A}^{2}\left(\dot{R}%
_A-HR_A\right) ,  \label{smd-1} \\
\dot{S}_{D} &=& \frac{1}{T}(1+\omega_D)\rho_{D} 4\pi R_{A}^{2}\left(\dot{R}%
_A-HR_A\right) .  \label{sdd-1}
\end{eqnarray}

Our next task is to determine the first order entropy variation of the
dynamical apparent horizon. This horizon is analogous to the event horizon
of a black hole and the temperature associated with it is given by \cite%
{Bak1,Hawking1,Jacobson1,Padmanabhan1} 
\begin{equation}  \label{tapp}
T_A=\frac{1}{2\pi R_A}.
\end{equation}%
\newline
As for the entropy, we shall employ the Barrow black hole entropy \cite%
{barrow}, with the standard horizon area given by $A_A=4\pi R_{A}^{2}$.
Thus, we obtain 
\begin{equation}  \label{sapp}
S_A=\gamma R_{A}^{\Delta +2},
\end{equation}
where $\gamma = (4\pi /A_0)^{1+\Delta/2}$. At this point, it is customary to
assume that in gravitational thermodynamics, the temperature of the
dynamical apparent horizon and that of the fluid inside are equal, otherwise
a temperature gradient might lead to nonequilibrium thermodynamics \cite%
{Padmanabhan2,Izquierdo1,Mimoso1}. Moreover, the energy flow might deform
the geometry \cite{Izquierdo1}. Now, differentiating equation (\ref{sapp}),
we get 
\begin{equation}
\dot{S}_{A}=(\Delta+2)\gamma R_{A}^{\Delta+1}\dot{R}_A.
\end{equation}
Finally, identifying $T$ in equations (\ref{smd-1}) and (\ref{sdd-1}) with $%
T_A$ in equation (\ref{tapp}), and adding equations (\ref{smd-1}), (\ref%
{sdd-1}), and (\ref{tapp}), we obtain the total entropy variation of the
thermodynamic system bounded by the dynamical apparent horizon \cite%
{Saridakis3}: 
\begin{eqnarray}
\dot{S}_{tot} &=& \dot{S}_{m}+\dot{S}_{D}+\dot{S}_{A}  \notag \\
&=& 8\pi^{2} R_{A}^{3}\left(\dot{R}_A-HR_A\right)\left[(1+\omega_{d})%
\rho_{D}+(1+\omega_{m})\rho_{m}\right]  \notag \\
&+& (\Delta +2)\gamma R_{A}^{\Delta+1}\dot{R}_A  \notag \\
&=& \frac{2\pi}{G}H^{-5}\dot{H}\left\{\dot{H}+H^2 \left[1-\frac{\gamma G}{%
2\pi}(\Delta +2)H^{-\Delta}\right]\right\} .  \label{stot-new}
\end{eqnarray}
In arriving at the last equality, we have used the relations $R_A=1/H$ and $%
\dot{R}_A=-\dot{H}R_{A}^{2}$. Now, taking out $H^2$ from within the braces
on the right hand side of equation (\ref{stot-new}) and noting that $\dot{H}/%
{H^2}=-1-q$, we obtain 
\begin{eqnarray}
\dot{S}_{tot} &=& \frac{2\pi}{G}H^{-3}\dot{H}\left[-q-\frac{\gamma G}{2\pi}%
(\Delta +2)H^{-\Delta}\right]  \notag \\
&=& -\frac{2\pi}{G}H^{-3}\dot{H}\left[\frac{%
(1-3b_{1}^{2})+(3b_{1}^{2}-3b_{2}^{2}-1-\Delta)\Omega_{D}}{%
2-(2-\Delta)\Omega_{D}}\right]  \notag \\
&-& \gamma (\Delta +2)H^{-(3+\Delta)}\dot{H} .  \label{stot-fin}
\end{eqnarray}
Let us now analyze equation (\ref{stot-fin}) mathematically. Observe that,
since the parameters $\gamma$ and $\Delta$ are nonnegative, so the second
term on the right hand side will be nonnegative if $\dot{H}<0$, i.e., when
the cosmic fluids respect the null-energy condition. This latter inequality
will also force the expression outside the square brackets in the first term
to remain positive. Therefore, in our proposed interacting BHDE model, the
GSL will be satisfied if 
\begin{equation}  \label{gsl-cond}
\xi = (1-3b_{1}^{2})+(3b_{1}^{2}-3b_{2}^{2}-1-\Delta)\Omega_{D} \geq 0
\end{equation}
due to the fact that the denominator inside the square bracket is always
positive. Note that the requirement (\ref{gsl-cond}) is sufficient and is by
no means necessary for the GSL to be satisfied. Two cases may arise:

\begin{enumerate}
\item[(a)] $1-3b_{1}^{2} \geq 0$: This gives $3b_{1}^{2} \leq 1$ which
implies that $3b_{1}^{2}-3b_{2}^{2}-1-\Delta \leq 0$. Thus, in this case,
GSL will be satisfied if 
\begin{equation}
\Omega_{D} \leq \frac{|3b_{1}^{2}-1|}{|3b_{1}^{2}-3b_{2}^{2}-1-\Delta|}.
\end{equation}

\item[(b)] $3b_{1}^{2}-1 > \mbox{max}\{0,|3b_{2}^{2}-\Delta|\}$: This
implies that GSL will be satisfied if 
\begin{equation}
\Omega_{D} > \frac{3b_{1}^{2}-1}{3b_{1}^{2}-3b_{2}^{2}-1-\Delta}.
\end{equation}
\end{enumerate}

It must, however, be noted that a third case is mathematically plausible
where $0 < 3b_{1}^{2}-1 < |3b_{2}^{2}-\Delta|$, but it turns out that these
inequalities lead us to an unphysical scenario: $\Omega_{D}<0$.\newline

If, on the other hand, $\dot{H}>0$, then we might safely deduce that the GSL
is violated if the iequality in (\ref{gsl-cond}) is satisfied. Again, this
is only a sufficient condition for the violation of GSL and is by no means
necessary.\newline

Therefore, the above analyses show that the violation of the GSL is a
possibility in our proposed interacting BHDE model depending on nature of
evolution of the Universe.

%%%%%%%%%%%%%%%%%%%%%%%%%%%%%%%%%%%%%%%%%%%%%%%%%%%%%%%%%%%%%%%%%%%%%

\section{Conclusions}

\label{conclusion} In this work, we have proposed a new interacting HDE
model which is based on the recently proposed Barrow entropy \cite{barrow},
which originates from the modification of the black-hole surface due to some
quantum-gravitational effects. As discussed in section \ref{sec2}, for $%
\Delta=0$, the BHDE coincides with the standard HDE, while for $0<\Delta<1$
it leads to a new and interesting cosmological scenario. In particular, we
have studied the evolution of a spatially flat FRW universe composed of
pressureless dark matter and BHDE that interact with each other through a
well-motivated interaction term given by equation (\ref{eq-ans}). By
considering the Hubble horizon as the infrared cut-off, we have then studied
the behavior of the density parameter of BHDE, the EoS parameter of BHDE and
the deceleration parameter, during the cosmic evolution. \newline

It has been found that the BHDE model exhibits a smooth transition from
early deceleration era ($q>0$) to the present acceleration ($q<0$) era of
the universe. Also, the value of this transition redshift is in well
accordance with the current cosmological observations \cite%
{zt1,zt2,zt3,zt4,zt5,zt6,zt7,zt8}. As discussed in section \ref{sec2}, it
has also been found that the evolution behaviors of $\omega _{D}$ and $%
\omega _{tot}$ are in good agreement with recent observations. The latter
behaviour it is justified by the main analysis on the asymptotic behavior
for evolution of the field equations, where the de Sitter universe is an
attractor for the cosmological model. Furthermore, the analytic solution of
the field equations was presented. That result it is essential because we
know that the numerical simulations describe actual solutions of the
dynamical system. \newline

Finally, we have studied the implications of gravity-thermodynamics in the
BHDE model by assuming the dynamical apparent horizon as the cosmological
boundary. The apparent horizon is endowed with Hawking temperature and
Barrow entropy defined in equations (\ref{tapp}) and (\ref{sapp})
respectively. In particular, we have examined the viability of the GSL.
After a careful mathematical analysis, we have found that there is a
possibility of conditional violation of the GSL based on how the Universe
undergoes evolution. More precisely, we have obtained certain constraints on
the density parameter $\Omega_D$ for which the GSL will be satisfied in the
case where $\dot{H}<0$, while, on the other hand, we have obtained a
condition for the violation of the GSL in the case where $\dot{H}>0$. One
must, however, note that these constraints are sufficient in nature and are
by no means necessary for the viability of the GSL.

%%%%%%%%%%%%%%%%%%%%%%%%%%%%%%%%%%%%%%%%%%%%%%%%%%%%%%%%%%%%%%%%%%%

\section{Acknowledgments}
The authors are thankful to an anonymous reviewer for valuable comments which have helped to improve the presentation of this work.

%%%%%%%%%%%%%%%%%%%%%%%%%%%%%%%%%%%%%%%%%%%%%%%%%%%%%%%%%%%%%%%%%%%%%%%%%%%%%%%%%%%%%


\begin{thebibliography}{999}
%\bibitem{} 
%%%%%%%%%%%%%%%%%%%%%%%%%%%%%%%%%%%%%%%%%%%%%%%%%%%%%%%%%%%%%%%%%%%%%
\bibitem{acc1} A. G. Riess et al., Astron. J. \textbf{116}, 1009 (1998).

\bibitem{acc2} S. Perlmutter et al., Astrophys. J. \textbf{517}, 565 (1999).

\bibitem{acc3} P.A.R. Ade et al., Astron. Astrophys. \textbf{571}, A16
(2014).

\bibitem{acc4} D.N. Spergel et al., Astrophys. J. Suppl. Ser. \textbf{148},
175 (2003).

\bibitem{acc5} M. Tegmark etal., Phys. Rev. D \textbf{69}, 103501 (2004). 
%%%%%%%%%%%%%%%%%%%%%%%%%%%%%%%%%%%%%%%%%%%%%%%%%%%%%%%%%%%%%%%%%%%%%%%%%%%%%%

\bibitem{de1} E. J. Copeland, M. Sami and S. Tsujikawa, Int. J. Mod. Phys.
D. \textbf{15}, 1753 (2006).

\bibitem{de2} L. Amendola and S. Tsujikawa, Dark Energy: Theory and
Observations, Cambridge University Press, Cambridge, UK (2010).

\bibitem{de3} K. Bamba, S. Capozziello, S. Nojiri and S. D. Odintsov,
Astrophys. Space Sci. \textbf{342}, 155 (2012). 
%%%%%%%%%%%%%%%%%%%%%%%%%%%%%%%%%%%%%%%%%%%%%%%%%%%%%%%%

\bibitem{hp1} G.'t Hooft, Salamfest 1993: 0284-296 [arXiv: gr-qc/9310026].

\bibitem{hp2} L. Susskind, J. Math. Phys. \textbf{36}, 6377 (1995).

\bibitem{hp3} W. Fischler and L. Susskind, arXiv: hep-th/9806039.

\bibitem{hp4} A. Cohen, D. Kaplan, A. Nelson, Phys. Rev. Lett. \textbf{82},
4971 (1999).

\bibitem{hp5} P. Horava and D. Minic, Phys. Rev. Lett. \textbf{85}, 1610
(2000).

\bibitem{hp6} R. Bousso, Rev. Mod. Phys. \textbf{74}, 825 (2002). 
%%%%%%%%%%%%%%%%%%%%%%%%%%%%%%%%%%%%%%%%%%%%%%%%%%%%%%%%%%

\bibitem{hde1} M. Li, Phys. Lett. B \textbf{603}, 1 (2004).

\bibitem{hde2} R. Horvat, Phys. Rev. D \textbf{70}, 087301 (2004).

\bibitem{hde3} Q. G. Huang and M. Li, JCAP \textbf{0408}, 013 (2004).

\bibitem{hde4} B. Wang, Y. g. Gong and E. Abdalla, Phys. Lett. B \textbf{624}%
, 141 (2005).

\bibitem{hde5} D. Pavon and W. Zimdahl, Phys. Lett. B \textbf{628}, 206
(2005).

\bibitem{hde6} H. Kim, H. W. Lee and Y. S. Myung, Phys. Lett. B \textbf{632}%
, 605 (2006).

\bibitem{hde7} S. Nojiri and S. D. Odintsov, Gen. Rel. Grav. \textbf{38},
1285 (2006).

\bibitem{hde8} M. R. Setare, Phys. Lett. B \textbf{642}, 1 (2006).

\bibitem{hde9} B. Wang, C. Y. Lin and E. Abdalla, Phys. Lett. B \textbf{637}%
, 357 (2006).

\bibitem{hde10} M. R. Setare and E. N. Saridakis, Phys. Lett. B \textbf{670}%
, 1 (2008).

\bibitem{hde11} S. Wang, Y. Wang and M. Li, Phys. Rept. \textbf{696}, 1
(2017).

\bibitem{hde12} E. N. Saridakis, K. Bamba, R. Myrzakulov and F. K.
Anagnostopoulos, JCAP \textbf{12}, 012 (2018).

\bibitem{hde13} S. Nojiri, S. D. Odintsov and E. N. Saridakis, Phys. Lett. B 
\textbf{797}, 134829 (2019).

\bibitem{hde14} R. G. Cai, Phys. Lett. B \textbf{657}, 228 (2007).

\bibitem{hde15} C. Q. Geng, Y. T. Hsu, J. R. Lu and L. Yin, Eur. Phys. J. C 
\textbf{80}, 21 (2020).

\bibitem{hde16} E. N. Saridakis, Phys. Lett. B \textbf{660}, 138 (2008).

\bibitem{hde17} M. R. Setare and E. C. Vagenas, Int. J. Mod. Phys. D \textbf{%
18}, 147 (2009).

\bibitem{hde18} Y. Gong and T. Li, Phys. Lett. B \textbf{683}, 241 (2010)

\bibitem{hde19} Y. G. Gong, Phys. Rev. D \textbf{70}, 064029 (2004).

\bibitem{hde20} L. N. Granda, A. Oliveros, Phys. Lett. B \textbf{669}, 275
(2008).

%%%%%%%%%%%%%%%%%%%%%%%%%%%%%%%%%%%%%%%%%%%%%%%%%%%%%%

\bibitem{hdeo1} X. Zhang and F. Q. Wu, Phys. Rev. D \textbf{72}, 043524
(2005).

\bibitem{hdeo2} C. Feng, B. Wang, Y. Gong and R. K. Su, JCAP \textbf{0709},
005 (2007).

\bibitem{hdeo3} M. Li, X. D. Li, S. Wang and X. Zhang, JCAP \textbf{0906},
036 (2009).

\bibitem{hdeo4} X. Zhang, Phys. Rev. D \textbf{79}, 103509 (2009).

\bibitem{hdeo5} J. Lu, E. N. Saridakis, M. R. Setare and L. Xu, JCAP \textbf{%
1003}, 031 (2010).

\bibitem{hdeo6} A. A. Mamon, Int. J. Mod. Phys. D \textbf{26}, 1750136
(2017).

\bibitem{hdeo7} E. Sadri, Eur. Phys. J. C \textbf{79}, 762 (2019).

\bibitem{hdeo8} R. D'Agostino, Phys. Rev. D \textbf{99}, 103524 (2019). 
%%%%%%%%%%%%%%%%%%%%%%%%%%%%%%%%%%%%%%%%%%%%%%%%%%%%%%%%

\bibitem{barrow} J. D. Barrow, Phys. Lett. B \textbf{808}, 135643 (2020). 
%%%%%%%%%%%%%%%%%%%%%%%%%%%%%%%%%%%%%%%%%%%%%%%%%%%%%%%%%%%%%%%%%%%%%%%%%%%%%%

\bibitem{Bekenstein1} J. D. Bekenstein, Lett. Nuovo Cim. \textbf{4}, 737
(1972).

\bibitem{Bekenstein2} J. D. Bekenstein, Phys. Rev. D \textbf{7}, 2333
(1973). 
%%%%%%%%%%%%%%%%%%%%%%%%%%%%%%%%%%%%%%%%%%%%%%%%%%%%%%%%%%%%%%%%%%%%%%%%%%%%%%%

\bibitem{Saridakis} E. N. Saridakis, Phys. Rev. D {\bf 102}, 123525 (2020). 
%%arXiv: 2005.04115.

\bibitem{Saridakis2} F. K. Anagnostopoulos, S. Basilakos, E. N. Saridakis,
Eur. Phys. J. C \textbf{80}, 826 (2020).

\bibitem{intreview} Y. L. Bolotin, A. Kostenko, O. A. Lemets and D. A.
Yerokhin, IJMPD, \textbf{24}, 1530007 (2015). 
%%%%%%%%%%%%%%%%%%%%%%%%%%%%%%%%%%%%%%%%%%%%%%%%%%%

\bibitem{h01} E. Di Valentino, A. Melchiorri and O. Mena, Phys. Rev. D 
\textbf{96}, 043503 (2017).

\bibitem{h02} S. Kumar and R. C. Nunes, Phys. Rev. D \textbf{96}, 103511
(2017) 
%%%%%%%%%%%%%%%%%%%%%%%%%%%%%%%%%%%%%%%%%%%%%%%%%%%%%%%%%%%%%%%%%%%%%%%%%%%%%%%%

\bibitem{te1} C. Tsallis, J. Statist. Phys. \textbf{52}, 479 (1988).

\bibitem{te2} C. Tsallis and L. J. L. Cirto, Eur. Phys. J. C \textbf{73},
2487 (2013).

\bibitem{new2} S. D. H. Hsu, Phys. Lett. B \textbf{594}, 13 (2004).

%\bibitem{int3} D. Pavon and W. Zimdahl, Phys. Lett. B \textbf{628}, 206 (2005).

\bibitem{new3} A. Sheykhi, Class. Quantum Grav. \textbf{27}, 025007 (2010).

%%%%%%%%%%%%%%%%%%%%%%%%%%%%%%%%%%%%%%%%%%%%%%%%%%%%%

\bibitem{int1} M. Quartin et al., J. Cosmol. Astropart. Phys, \textbf{05},
007 (2008).

\bibitem{int2} D. Pavon and B. Wang, Gen. Rel. Grav. \textbf{41}, 1 (2009).

\bibitem{int4} C. G. Boehmer, G. Caldera-Cabral, R. Lazkoz, and R. Maartens,
Phys. Rev. D \textbf{78}, 023505 (2008).

\bibitem{int5} G. Caldera-Cabral, R. Maartens, and L. A. Urena-Lopez, Phys.
Rev. D \textbf{79}, 063518 (2009).

\bibitem{int6} A. A. Mamon, A. H. Ziaie and K. Bamba, Eur. Phys.  J. C \textbf{80}, 974 (2020). 
%%arXiv: 2004.01593. 
%%%%%%%%%%%%%%%%%%%%%%%%%%%%%%%%%%%%%%%%%%%

\bibitem{intobs} L. P. Chimento, Phys. Rev. D \textbf{81}, 043525 (2010). 
%%%%%%%

\bibitem{inan1} G.\ Papagiannopoulos, P. Tsiapi, S. Basilakos and A.
Paliathanasis, EPJC \textbf{80}, 55 (2020).

\bibitem{inan2} S. Pan, W. Yang and A. Paliathanasis, MNRAS \textbf{493},
3114 (2020).

\bibitem{sp1} W. Yang, S. Pan, E. Di Valentino, R.C. Nunes, S. Vagnozzi and
D.F. Mota, JCAP \textbf{09}, 019 (2018).

\bibitem{sp2} W. Yang, A. Mukherjee, E. Di Valentino and S. Pan, Phys. Rev.
D \textbf{98}, 123527 (2018).

\bibitem{sp3} W. Yang, S. Pan and A. Paliathanasis, MNRAS \textbf{482}, 1007
(2019).

\bibitem{epo1} T. Gonzalez, G.\ Leon and I. Quiros, Class. Quantum Grav. 
\textbf{30}, 135001 (2013).

\bibitem{epo2} G. Leon, Class. Quantum Grav. \textbf{26}, 035008 (2009).

\bibitem{epo3} A. Giacomini, G.\ Leon, A. Paliathanasis and S. Pan, Eur.
Phys. J. C \textbf{80}, 1 (2020).

\bibitem{epo4} R. Lazkoz, G. Leon and I. Quiros, Phys. Lett. B \textbf{649},
103 (2007).

%%%%%%%%%%%%%%%%%%%%%%%%%%%%%%%%%

\bibitem{zt1} O. Farooq, B. Ratra, Astrophys. J. \textbf{766}, L7 (2013).

\bibitem{zt2} O. Farooq, F. R. Madiyar, S. Crandall, B. Ratra, ApJ \textbf{%
835}, 26 (2017).

\bibitem{zt3} A. A. Mamon, K. Bamba, S. Das, Eur. Phys. J. C \textbf{77}, 29
(2017).

\bibitem{zt4} A. A. Mamon and K. Bamba, Eur. Phys. J. C \textbf{78}, 862
(2018).

\bibitem{zt5} J. Magana et al., J. Cosmol. Astropart. Phys. \textbf{10}, 017
(2014).

\bibitem{zt6} A. A. Mamon, S. Das, Int. J. Mod. Phys. D. \textbf{25},
1650032 (2016).

\bibitem{zt7} A. A. Mamon, S. Das, Eur. Phys. J. C \textbf{77}, 495 (2017).

\bibitem{zt8} A. A. Mamon, Mod. Phys. Lett. A \textbf{33}, 1850056 (2018).

%%%%%%%%%%%%%%%%%%%%%%%%%%%%%%%%%%%%%%%%%%%%%%%%%%%%%%%%%%

\bibitem{Abl1} M.J. Ablowitz, A. Ramani and H. Segur, \ Lettere al Nuovo
Cimento \textbf{23} 333 (1978).

\bibitem{Abl2} M.J. Ablowitz, A. Ramani and H. Segur, J. Math. Phys. \textbf{%
21} 715 (1980).

\bibitem{Abl3} M.J. Ablowitz, A. Ramani and H. Segur, J. Math. Phys. \textbf{%
21} 1006 (1980).

\bibitem{miritzis} J. Miritzis, P.G.L. Leach and S.\ Cotsakis, Grav. Cosmol. 
\textbf{6} 282 (2000).

\bibitem{cots1} S. Cotsakis and P.G.L. Leach, J. Phys A \textbf{27}, 1625
(1994).

\bibitem{cots2} P.G.L.\ Leach, S. Cotsakis and J. Miritzis, Grav. Cosmol. 
\textbf{7}, 311 (2000).

\bibitem{bun2} T. Bountis and L. Drossos, On The Non-Integrability of the
Mixmaster Universe Model. In: Sim\'{o} C. (eds) Hamiltonian Systems with
Three or More Degrees of Freedom. NATO ASI Series (Series C:Mathematical and
Physical Sciences), vol 533. Springer, Dordrecht (1999).

\bibitem{cots} S. Cotsakis, J. Demaret, Y. De Rop and L. Querella, Phys.
Rev. D \textbf{48}, 4595 (1993).

\bibitem{sinFR} A. Paliathanasis and P.G.L. Leach, Phys.\ Lett. A \textbf{380%
}, 2815 (2016).

\bibitem{sinFR2} A.\ Paliathanasis, EPJC \textbf{77}, 438 (2017).

\bibitem{sinFT} A. Paliathanasis, J.D. Barrow and P.G.L. Leach, Phys. Rev. D 
\textbf{94}, 023525 (2016).

\bibitem{sinr02} S. Cotsakis, S. Kadry, G. Kolionis and A. Tsokaros, Phys.
lett. B \textbf{755}, 387 (2016). 
%%%%%%%%%%%%%%%%%%%%%%%%%%%%%%%%%%%%%%%%%%%%%%%%%%%%
%%%%%%%%%%%%%%%%%%%%%%%%%%%%%%%%%%%%%%%%%%%%%%%%%%%%%%%%%%%%%%%%%%%%%%%%%%%%%%%%%

\bibitem{Wang1} B. Wang, Y. Gong, and E. Abdalla, Phys. Rev. D \textbf{74},
083520 (2006).

\bibitem{Saha1} S. Saha and S. Chakraborty, Phys. Lett. B \textbf{717}, 319
(2012).

\bibitem{Saha2} S. Saha and S. Chakraborty, Phys. Rev. D \textbf{89}, 043512
(2014).

\bibitem{Chakraborty1} S. Chakraborty and S. Saha, Mod. Phys. Lett. A 
\textbf{30}, 1550024 (2015).

\bibitem{Saha3} S. Saha, Int. J. Mod. Phys. A \textbf{34}, 1950193 (2019).

\bibitem{Mamon1} A. A. Mamon and S. Saha, Int. J. Mod. Phys. D, \textbf{29}, 2050097 (2020).
%%DOI: 10.1142/S0218271820500972; arXiv: 2003.08328. 
%%International Journal of Modern Physics DVol. 29, No. 15, 2050097 (2020) 
%%%%%%%%%%%%%%%%%%%%%%%%%%%%%%%%%%%%%%%%%%%%%%%%%%%%%%%%%

\bibitem{Hayward1} S. A. Hayward, Class. Quant. Grav. \textbf{15}, 3147
(1998).

\bibitem{Hayward2} S. A. Hayward, S. Mukohyama and M. Ashworth, Phys. Lett.
A \textbf{256}, 347 (1999).

\bibitem{Bak1} D. Bak and S. J. Rey, Class. Quant. Grav. \textbf{17}, L83
(2000).

\bibitem{Saridakis3} E. N. Saridakis and  S. Basilakos, arXiv: 2005.08258 (2020).

\bibitem{Hawking1} S. W. Hawking, Commun. Math. Phys. \textbf{43}, 199
(1975).

\bibitem{Jacobson1} T. Jacobson, Phys. Rev. Lett. \textbf{75}, 1260 (1995).

\bibitem{Padmanabhan1} T. Padmanabhan, Class. Quant. Grav. \textbf{19}, 5387
(2002).

\bibitem{Padmanabhan2} T. Padmanabhan, Rept. Prog. Phys. \textbf{73}, 046901
(2010).

\bibitem{Izquierdo1} G. Izquierdo and D. Pav\'{o}n, Phys. Lett. B \textbf{633%
}, 420 (2006).

\bibitem{Mimoso1} J P. Mimoso and D. Pav\'{o}n, Phys. Rev. D \textbf{94},
103507 (2016). %%%%%%%%%%%%%%%%%%%%%%%%%%%%%%%%
%%%%%%%%%%%%%%%%%%%%%%%%%%%%%%%%%%%%%%%%%%%%%%%%%%%%%%%%%%
\end{thebibliography}
\end{document}